\documentclass[reprint,aps]{revtex4-1}
\usepackage{amsmath}
\usepackage{amssymb}
\usepackage{bm}
\usepackage{braket}
\usepackage{color}
\usepackage[varg]{txfonts}
\usepackage{varwidth}
\usepackage{dcolumn}
\usepackage[breaklinks,colorlinks=true,linkcolor=blue,urlcolor=cyan,citecolor=blue]{hyperref}
\usepackage{setspace}
\usepackage{here}
\usepackage{graphicx}

\newcommand{\bettershortstack}[4][c]{
 \renewcommand{\arraystretch}{#2}
 \begin{tabular}[b]{@{}#1@{}}
 #4
 \end{tabular}
 \renewcommand{\arraystretch}{#3}
}

\begin{document}

\title{Crystal-field Paschen-Back effect on ruby in ultrahigh magnetic fields}

\author{Masaki Gen}
 \email{gen@issp.u-tokyo.ac.jp}
 \affiliation{Institute for Solid State Physics, University of Tokyo, Kashiwa, Chiba 277-8581, Japan}

\author{Tomoki Kanda} 
 \affiliation{Institute for Solid State Physics, University of Tokyo, Kashiwa, Chiba 277-8581, Japan}

\author{Takashi Shitaokoshi} 
 \affiliation{Institute for Solid State Physics, University of Tokyo, Kashiwa, Chiba 277-8581, Japan}

\author{Yoshimitsu Kohama} 
 \affiliation{Institute for Solid State Physics, University of Tokyo, Kashiwa, Chiba 277-8581, Japan}

\author{Toshihiro Nomura}
 \email{t.nomura@issp.u-tokyo.ac.jp}
 \affiliation{Institute for Solid State Physics, University of Tokyo, Kashiwa, Chiba 277-8581, Japan}

\begin{abstract}

Zeeman spectra of the R lines of ruby (Cr$^{3+}$: $\alpha$-Al$_{2}$O$_{3}$) were studied in ultrahigh magnetic fields up to 230~T by magneto-photoluminescence measurements.
The observed Zeeman patterns exhibit nonlinear behaviors above 100~T, evidencing the breakdown of the previously reported Paschen-Back effect for ${\bf B} \perp c$ geometry.
We adopted the crystal-field multiplet theory including the cubic crystal field (${\mathcal H}_{\rm cubic}$), the trigonal crystal field (${\mathcal H}_{\rm trig}$), the spin-orbit interaction (${\mathcal H}_{\rm SO}$), and the Zeeman interaction (${\mathcal H}_{\rm Z}$).
It is found that the nonlinear splitting of the R lines is owing to the hybridization between the $^{2}E$ and $^{2}T_{1}$ states, which leads to the quantization of these Zeeman levels with the orbital angular momentum.
Our results suggest that the exquisite energy balance among ${\mathcal H}_{\rm cubic}$, ${\mathcal H}_{\rm trig}$, ${\mathcal H}_{\rm SO}$, and ${\mathcal H}_{\rm Z}$ realized in ruby offers a unique opportunity to observe the onset of the {\it crystal-field} Paschen-Back effect toward the high-field extreme.

\end{abstract}

\date{\today}
\maketitle

\section{\label{sec:level1}INTRODUCTION}
\vspace{-0.2cm}

Zeeman effect is categorized into anomalous Zeeman (AZ) effect at the weak-field limit and normal Zeeman (NZ) effect at the high-field limit.
When atoms are located in weak magnetic fields, the energy levels split nonlinearly due to the competition between the external magnetic field and the hyperfine or the spin-orbit interactions.
Under high magnetic fields where the Zeeman energy far exceeds those interactions, on the other hand, the energy splitting is asymptotically quantized by $\mu_{\rm B}B$, where $\mu_{\rm B}$ is the Bohr magneton and $B$ is the magnetic field.
Historically, the crossover from the AZ effect to the NZ effect, the so-called Paschen-Back (PB) effect \cite{1921_Pas}, has been observed in various atoms \cite{1914_Ken, 1938_Kap, 1985_Win, 1988_Win, 1982_Hor, 2014_Sar, 2017_Sar} and molecules \cite{1949_Jen, 2006_Ber}.
For example, the D lines of the sodium atom exhibit the hyperfine PB effect around 30~mT \cite{1985_Win, 1988_Win} and the spin-orbit PB effect around 50~T \cite{1982_Hor}.
Through this process, the good quantum number changes from $M_{F}$ for the total angular momentum of an atom to $M_{I}$ and $M_{J}$ for that of a nucleus and electrons, respectively, and then $M_{J}$ are further decoupled to $M_{L}$ and $M_{S}$ for the orbital and spin angular momenta, respectively.

\begin{figure}[b]
\centering
\includegraphics[width=\columnwidth]{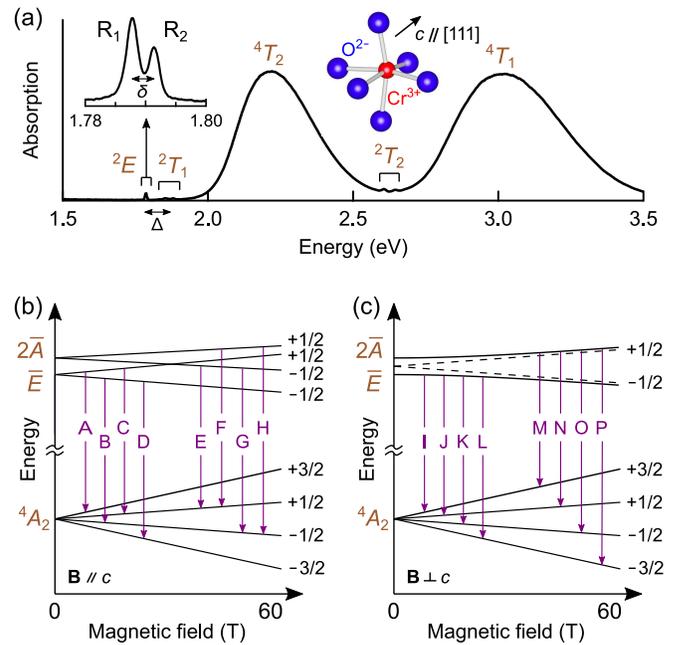}
\caption{(a) Absorption spectrum of ruby in visible region and the local structure of CrO$_{6}^{9-}$ with the $C_{3}$ symmetry. The corresponding excited state from the ground state $^{4}A_{2}$ is denoted for each absorption peak. (b)(c) Energy diagrams of the Zeeman splitting of the R lines for (b) ${\bf B} \parallel c$ and (c) ${\bf B} \perp c$ geometries. A small initial splitting of 0.05~meV in the $^{4}A_{2}$ state is neglected. Optically allowed transitions for ${\bf E} \perp c$ are denoted by A, B, ..., P. In the PB region for ${\bf B} \perp c$, the excited levels asymptotically approach the dashed straight lines.}
\label{Introduction}
\end{figure}

The PB effect can also be observed in solid states \cite{1965_Woo, 1971_Kos, 1958_Sug_2, 1963_Aoy, 1979_Hor, 2008_Mil}.
The most well-known example is the R lines of ruby \cite{1958_Sug_2, 1963_Aoy, 1979_Hor, 2008_Mil}.
Many spectroscopic works on ruby have been done for more than half century, motivated by the scientific interests \cite{1958_Sug_2, 1963_Aoy, 1979_Hor, 2008_Mil, 1958_Tan, 1958_Sug_1, 1961_Sug, 1970_Mac, 1955_Man, 1982_Kido, 2009_Wat, 2010_Mil, 2011_Wat, 2012_Ara, 2018_Hun} as well as its applications to a solid-state laser and pressure gauge \cite{1960_Mai_1, 1960_Mai_2, 1972_For, 1986_Mao, 1991_Gup, 1992_Rag}.
The optical transitions of ruby stem from the Cr$^{3+}$ impurities in $\alpha$-Al$_{2}$O$_{3}$, where Cr$^{3+}$ ions are subjected to the cubic crystal field of the octahedrally coordinated O$^{2-}$ ions.
Furthermore, the repulsion between neighboring cations causes the slight lattice distortion, lowering the symmetry to trigonal $C_{3}$.
As shown in Fig. \ref{Introduction}(a), the absorption spectrum of ruby in visible region consists of two broad bands and three groups of sharp lines.
In the notation of the cubic symmetry, the R lines correspond to an electric dipole transition from the ground state $^{4}A_{2}$ to the first excited state $^{2}E$, appearing around 1.79~eV (694~nm).
Due to the interplay of the trigonal crystal field and the spin-orbit interaction, the $^{2}E$ state further splits into two Kramers doublets $\overline E$ and 2$\overline A$ (R$_{1}$ and R$_{2}$ lines) with the energy gap of $\delta=3.6$~meV.
The observed Zeeman patterns of the R lines up to 60~T agree well with the theory based on the effective Hamiltonian for the $^{2}E$ state, which are shown in Figs. \ref{Introduction}(b) and \ref{Introduction}(c) \cite{1963_Aoy, 1979_Hor, 2008_Mil, 1958_Tan, 1958_Sug_1}.
Here, eight optically allowed transitions for ${\bf E} \perp c$ are noted for each ${\bf B} \parallel c$ and ${\bf B} \perp c$ geometry, where $c$ axis is taken as the trigonal axis.
For ${\bf B} \perp c$, the tendency of the PB effect is observed toward 60~T, where the Zeeman energy is large enough compared to the initial splitting ($g\mu_{\rm B}B\approx 2\delta$, where $g=2$) \cite{1979_Hor, 2008_Mil}.

This PB effect in ruby can be understood as such, that the Cr$^{3+}$ spins initially oriented along the $c$ axis are gradually quantized along the applied magnetic field ${\bf B} \perp c$.
Unlike the case of isolated atoms, the Zeeman levels of the R lines become quantized only with $M_{S}$ in this PB region, while the orbital angular momentum of the $^{2}E$ state is still quenched in crystal.
Hence, one can naively expect further quantization of these levels with $M_{L}$ toward the high-field limit.

In this work, we observe the Zeeman spectra of the R lines of ruby in ultrahigh magnetic fields up to 230~T.
For both ${\bf B} \parallel c$ and ${\bf B} \perp c$ geometries, the Zeeman patterns exhibit nonlinear splitting above 100~T, already beyond the previous theory \cite{1958_Tan, 1958_Sug_1}.
The observed Zeeman patterns are analyzed by the standard crystal-field multiplet theory including the cubic crystal field, the trigonal crystal field, the spin-orbit interaction, and the Zeeman interaction \cite{1958_Tan, 1958_Sug_1, 1961_Sug, 1943_Rac, 1954_Tan_1, 1954_Tan_2, 1960_Tan}.
The hybridizations between the first and the second excited states, $^{2}E$ and $^{2}T_{1}$, are found to be responsible for the AZ effect above~100 T.
Our observation would be regarded as the onset of the {\it crystal-field} PB effect, originating from the competition between the Zeeman energy and the crystal-field splitting.

This paper is organized as follows.
In Sec.~\ref{sec:level2}, we present the experimental and theoretical methods.
In Sec.~\ref{sec:level3}, we show the observed and calculated results on the Zeeman patterns of the R lines.
In Sec.~\ref{sec:level4}, we discuss the wave functions of the $^{2}E$ states in magnetic fields.
We also refer to the Zeeman patterns at the high-field extreme.
In Sec.~\ref{sec:level5}, we end the paper with the conclusive remarks.

\vspace{-0.3cm}
\section{\label{sec:level2}METHODS}
\vspace{-0.2cm}

\subsection{Experimental setup}
\vspace{-0.2cm}

\begin{figure}[t]
\centering
\includegraphics[width=\linewidth]{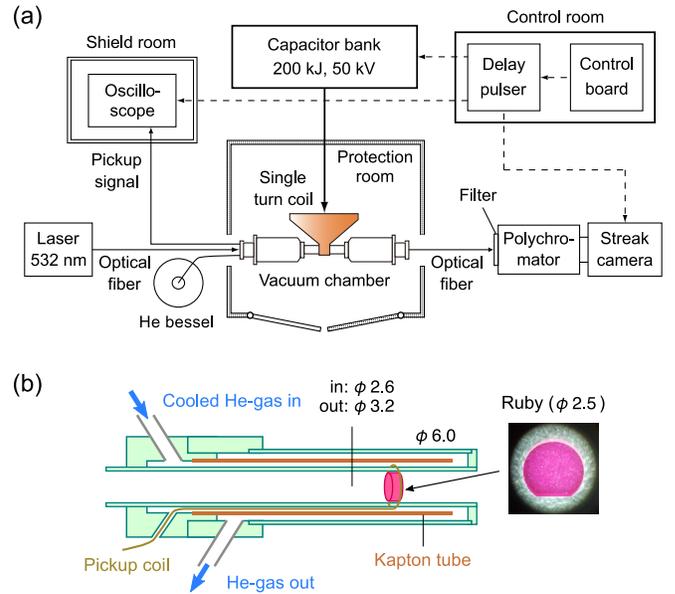}
\caption{(a) Block diagram of the experimental setup for the magneto-PL measurements using the STC system. The dashed lines show the trigger signals. (b) Schematic view of the $^{4}$He-flow-type cryostat and the photo of the ruby single crystal used in this work.}
\label{Setup}
\end{figure}

Figure \ref{Setup}(a) shows the block diagram of the experimental setup for the magneto-photoluminescence (PL) measurements.
The ultrahigh magnetic fields were generated by a horizontal single-turn-coil (HSTC) megagauss generator, equipped with 200~kJ fast capacitor banks.
The magnetic fields were measured by a calibrated pickup-coil with the error of $\pm 2$~\%.
Disk-shaped ruby single crystals with 2.5~mm diameter and 1.0~mm thickness (SHINKOSHA Co., Ltd.) were used.
The concentration of the Cr$^{3+}$ ion was 0.70~wt$\%$, which is low enough to neglect the effect of the exchange interaction.
The PL spectra were measured using a high-speed streak camera.
The 532~nm laser was used for the excitation $^{4}A_{2} {\rightarrow} ^{4}T_{2}$.
The incident light was guided by the 2.0-mm-diameter optical fiber and radiated the sample located at the center of the STC.
The emission light was collected by the 0.8-mm-diameter optical fiber and guided to the polychromator equipped with the green filter, then resolved by the streak camera.
In our setup, the incident light is polarized as ${\bf E} \perp c$ for ${\bf B} \parallel c$, while both ${\bf E} \perp c$ and ${\bf E} \parallel c$ components are mixed for ${\bf B} \perp c$.

The sample was cooled down by using $^{4}$He-flow-type cryostats, as shown in Fig. \ref{Setup}(b).
The sample and the optical fibers were inserted inside the cryostat.
A chromel/constantan thermocouple was tightly inserted to the half-moon-shaped gap between the sample and the cryostat.
The measurements were performed around 200~K.
In this temperature range, electrons are thermally distributed within the $^{2}E$ levels and most of the optically allowed transitions are observed as relatively sharp peaks.
Exceptionally, the measurements above 190~T were performed without using the cryostat due to the limited space inside the STC.
For the best signal-to-noise ratio, we performed several cycles of measurements with the different maximum fields $B_{\rm max}$ and integrated the spectra recorded near $B_{\rm max}$ within 1~\% for each pulse.

\begin{figure*}[t]
\centering
\includegraphics[width=\linewidth]{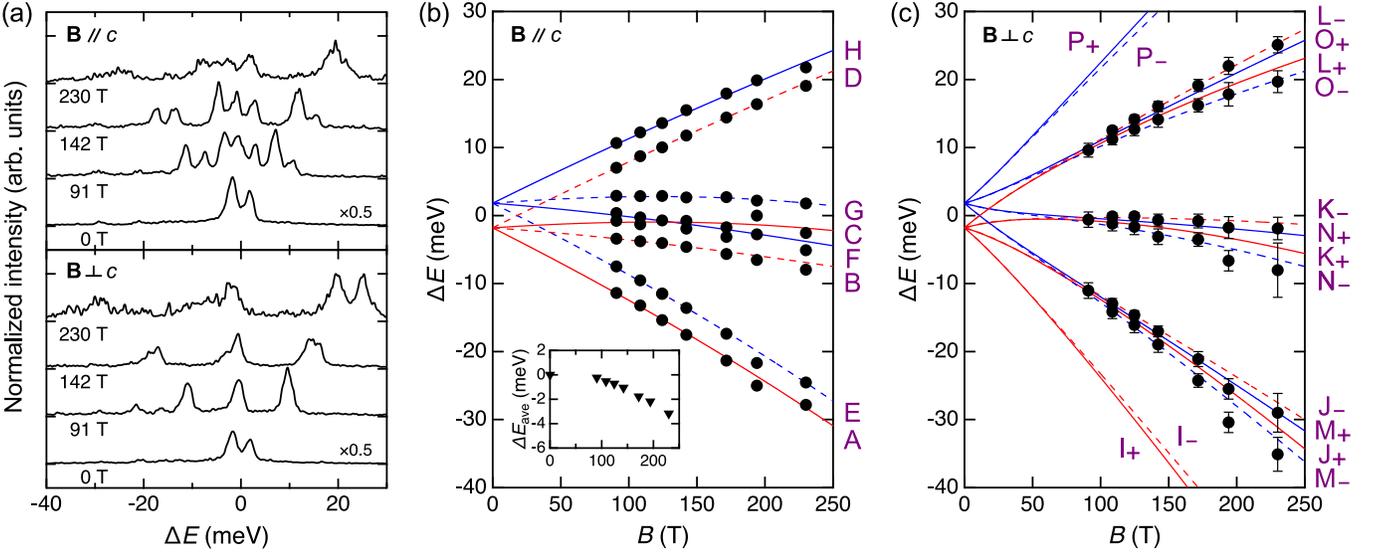}
\caption{(a) PL spectra of the R lines up to 230~T for ${\bf B} \parallel c$ (${\bf B} \perp c$) in the upper (lower) panel. The spectra were recorded around 200~K except for the 230~T data. (b)(c) Zeeman patterns of the R lines for (b) ${\bf B} \parallel c$ and (c) ${\bf B} \perp c$ geometries. Filled circles show the experimental peak positions. Red and blue lines show the theoretical curves obtained from the crystal-field multiplet theory (see text for details). Inset in (b) presents the average of the peak shifts with the magnetic field, reflecting the change in the mean energy of the four Zeeman levels splitting from the $^{2}E$ states. Error bars in (c) are taken from the full width at half maximum (FWHM) of the fitted Lorentzian functions.}
\label{Experiment}
\end{figure*}

\vspace{-0.3cm}
\subsection{Theoretical approach}
\vspace{-0.2cm}

So far, a simple effective Hamiltonian for the $^{2}E$ state has been considered to explain the Zeeman patterns of the R lines \cite{1963_Aoy, 1979_Hor, 2008_Mil, 1958_Tan, 1958_Sug_1}.
For ${\bf B} \parallel c$, the Zeeman patterns of the R$_{1}$ and R$_{2}$ lines in the low-field region can be respectively discribed by $E_{1\pm}(B)=-\frac{\delta}{2}\pm\frac{1}{2}(g_{0}+2g_{1\parallel})\mu_{\rm B}B$ and $E_{2\pm}(B)=\frac{\delta}{2}\pm\frac{1}{2}(g_{0}-2g_{2\parallel})\mu_{\rm B}B$ [see Fig. \ref{Introduction}(b)].
Here, $g_{0}=1.98$ \cite{1955_Man} is the $g$-factor of the ground state, and $g_{1\parallel}=0.23$ and $g_{2\parallel}=0.26$ \cite{2008_Mil} are the $g$-shifts which are mainly caused by the third-order interactions between the $^{2}E$ and the upper excited states, $^{2}T_{1}$ and $^{2}T_{2}$, through the trigonal crystal field and the orbital angular momentum along the $c$ axis ($L_{z}$) \cite{1958_Sug_1}.
For ${\bf B} \perp c$, in contrast, the Zeeman patterns are described by the quadratic relations without lifting the degeneracies of the Kramers doublets as $E_{1\pm}(B)=-(1/2)\sqrt{\delta^{2}+(g_{0}\mu_{\rm B}B)^2}$ and $E_{2\pm}(B)=(1/2)\sqrt{\delta^{2}+(g_{0}\mu_{\rm B}B)^2}$ [see Fig. \ref{Introduction}(c)].
As is evident from these formula, this model assumes that the center of the four energy levels remains constant in magnetic fields.

In this work, we adopted a Hamiltonian comprised of all the 120 bases in the $3d^{3}$ multiplet, which is more general treatment than the above perturbative one.
In this Hamiltonian, the trigonal crystal field (${\mathcal H}_{\rm trig}$), the spin-orbit interaction (${\mathcal H}_{\rm SO}$), and the Zeeman interaction (${\mathcal H}_{\rm Z}$) were involved in together with the cubic crystal field (${\mathcal H}_{\rm cubic}$).
The bases are expressed as $\ket{(\alpha S\Gamma) M_{S}\gamma}$, where $\alpha$ is the electronic configuration, $M_{S}$ the spin quantum number in the spin-$S$ state, and $\gamma$ the orbital function in the cubic irreducible representation $\Gamma$.
We take $M_{S}$ and $\gamma$ quantized along the trigonal $c$ axis ($u_{\pm}$ for $E$, $a_{\pm}$ and $a_{0}$ for $T_{1}$, and $x_{\pm}$ and $x_{0}$ for $T_{2}$) \cite{1958_Tan, 1958_Sug_1}.
In the following, the notation $(\alpha S\Gamma)$ is omitted if it is evident from the context.

Several empirical parameters were introduced in the multiplet Hamiltonian for numerical diagonalization: the cubic crystal-field strength $10Dq$, Racah parameters $B$ and $C$ \cite{1943_Rac}, the trigonal crystal fields $K$ and $K'$ defined as $K \equiv \bra{(t_{2}){+\frac{1}{2}x_{+}}}{\mathcal H}_{\rm trig}\ket{(t_{2}){+\frac{1}{2}x_{+}}}$ and $K' \equiv -(1/\sqrt{2})\bra{(t_{2}){+\frac{1}{2}x_{+}}}{\mathcal H}_{\rm trig}\ket{(e){+\frac{1}{2}u_{+}}}$, the spin-orbit interactions $\zeta$ and $\zeta'$ defined as $\zeta \equiv -2\bra{(t_{2}){+\frac{1}{2}x_{+}}}{\mathcal H}_{\rm SO}\ket{(t_{2}){+\frac{1}{2}x_{+}}}$ and $\zeta' \equiv -\sqrt{2}\bra{(t_{2}){+\frac{1}{2}x_{+}}}{\mathcal H}_{\rm SO}\ket{(e){+\frac{1}{2}u_{+}}}$, and the orbital reduction factors $k$ and $k'$ defined as $k \equiv -\bra{(t_{2}){+\frac{1}{2}x_{+}}}L_z\ket{(t_{2}){+\frac{1}{2}x_{+}}}$ and $k' \equiv -(1/\sqrt{2})\bra{(t_{2}){+\frac{1}{2}x_{+}}}L_z\ket{(e){+\frac{1}{2}u_{+}}}$, which reflect the Cr-O bond covalency.
Note that $K$, $K'$, $\zeta$, $\zeta'$, $k$, and $k'$ are the matrix elements between one-electron states, and $K=K'$, $\zeta=\zeta'$, and $k=k'=1$ hold for the free ion.
Similar theoretical approach focusing on the low-field limit was attempted previously \cite{1961_Sug, 1970_Mac}, but the discrepancies between experimental and theoretical values of $\delta$ and $g$-factor were relatively large.
Hence, although no perfect quantitative match seems to be achieved with any set of parameters, reexamination of the appropriate values was required in this work.
We chose the parameters in ${\mathcal H}_{\rm cubic}$ as $10Dq=2.320$~eV, $B=0.071$~eV, and $C=0.429$~eV following the latest analysis in Ref. \cite {2018_Hun}, which succeeded in reproducing the spectrum of ruby at zero field in a wide energy range including the UV region.
Then, we searched for the combination of the values of $K$, $K'$, $\zeta$, $\zeta'$, $k$, and $k'$ which simultaneously satisfy (i) the initial splitting $\delta=3.6$~meV, (ii) the $g$-values of the R$_{1}$ and R$_{2}$ lines for ${\bf B} \parallel c$ in the low-field limit, $g_{0}+2g_{1\parallel}=2.44$ and $g_{0}-2g_{2\parallel}=1.46$ \cite{2008_Mil}, and (iii) our new experimental results in the high-field region.
Here, we imposed additional constraints of $K<0$, $\zeta>0$, $0<K'/K<1$, $0.8<\zeta'/\zeta<1$, and $k'/k=\zeta'/\zeta$.

\vspace{-0.3cm}
\section{\label{sec:level3}RESULTS}
\vspace{-0.2cm}

\subsection{Experimental Zeeman patterns}
\vspace{-0.2cm}

The evolutions of the Zeeman spectra of the R lines are shown in Fig. \ref{Experiment}(a) (For all the data sets, see Supplemental Material \cite{Supple}).
The peak positions as a function of magnetic field for ${\bf B} \parallel c$ and ${\bf B} \perp c$ are plotted in Figs. \ref{Experiment}(b) and \ref{Experiment}(c), respectively, which are extracted by multi-Lorentzian fits \cite{Supple}.
The energy shift $\Delta E$ is measured from the center of the R$_{1}$ and R$_{2}$ lines at zero field.
For ${\bf B} \parallel c$, all of the eight optically allowed lines, corresponding to A--H in Fig. \ref{Introduction}(b), are clearly observed.
Remarkably, the Zeeman patterns deviate from the linear field dependence to the lower energy side as the magnetic field increases, resulting in the shift of the average peak position as shown in the inset of Fig. \ref{Experiment}(b).
For ${\bf B} \perp c$, three distinct peaks are observed at 91~T, which can be understood that six lines, corresponding to J--O in Fig. \ref{Introduction}(c), merge into three lines in the previously reported PB region.
Note that the rest two lines, I and P, are hardly observable due to the little transition probability under high magnetic fields \cite{1963_Aoy}.
Surprisingly, those three peaks split again into (at least) six peaks above 100~T, contradicting the concept of the previously reported PB effect.
The tendency of the peak shifts to the lower energy side is also seen in the case of ${\bf B} \perp c$.

\begin{figure}[t]
\centering
\includegraphics[width=\columnwidth]{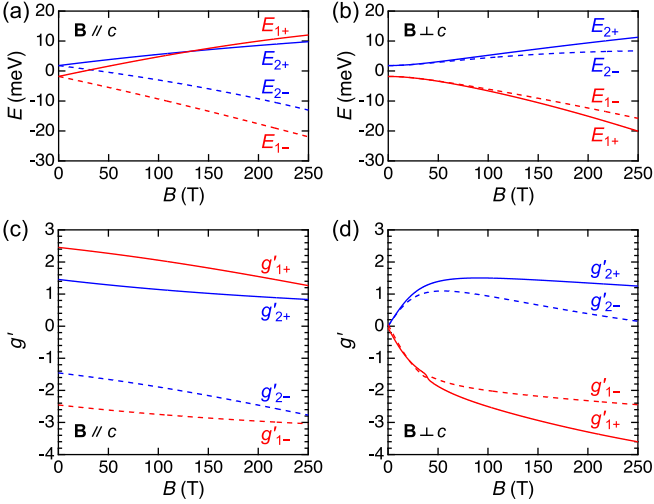}
\caption{(a)(b) Calculated energy diagrams of the four Zeeman levels splitting from the $^{2}E$ states with the magnetic field up to 250~T for (a) ${\bf B} \parallel c$ and (b) ${\bf B} \perp c$ geometries. (c)(d) The corresponding field-derivatives of the energy for (c) ${\bf B} \parallel c$ and (d) ${\bf B} \perp c$ geometries.}
\label{Calculation}
\end{figure}

\vspace{-0.3cm}
\subsection{Theoretical energy diagrams}
\vspace{-0.2cm}

The calculated energy diagrams of the four Zeeman levels splitting from the $^{2}E$ states are shown up to 250~T in Figs. \ref{Calculation}(a) and \ref{Calculation}(b).
In the crystal-field multiplet Hamiltonian, the parameters are chosen as $K=-0.037$~eV, $K'=-0.013$~eV, $\zeta=0.025$~eV, $k=0.72$, and $k'/k=\zeta'/\zeta=0.86$.
Here, $K$ and $K'$ are a bit smaller compared to those in the previous works, whereas $\zeta$ and $\zeta'$ the opposite \cite{1961_Sug, 1970_Mac, 2018_Hun}.
It can be seen that all of the four curves for each geometry show upwardly convex magnetic-field dependence above~100 T.
These trends can be clearly captured as the decrease in the field-derivatives of the energy $g'_{r}\equiv \frac{2}{\mu_{\rm B}}\frac{dE_{r}}{dB}$ ($r=1\pm, 2\pm$) as shown in Figs. \ref{Calculation}(c) and \ref{Calculation}(d).
Furthermore, for ${\bf B} \perp c$, the splitting of each Kramers doublet starts to be seen around 100~T and becomes larger toward higher fields as shown in Fig. \ref{Calculation}(b).
The ground state $^{4}A_{2}$ is found to show the nearly linear Zeeman splitting up to 250~T (not shown).
From the above, the theoretical Zeeman patterns of the R lines are obtained as shown in Figs. \ref{Experiment}(b) and \ref{Experiment}(c), which agree well with the experimental peak plots for both geometries.
This indicates that the semiempirical crystal-field multiplet Hamiltonian taking all the bases in the $3d^{3}$ state can account for the Zeeman spectra of the R lines of ruby even in the megagauss region.

As shown in Fig. \ref{Experiment}(c), our calculation predicts that each of the three merged peaks around 100~T splits into four individual peaks beyond the previously reported PB region for ${\bf B} \perp c$.
All of them are optically allowed, suggesting that the observed PL spectra above 100~T are composed of 12 peaks, while not all lines are well resolved due to their overlapping.
Those spectra are tentatively fitted by six peaks with relatively large errors because the peak assignments with the 12-peak fit are challenging within our experimental accuracy.
The detailed peak assignments based on the discussion on the peak intensities are found in Appendix A.

\begin{figure}[b]
\centering
\includegraphics[width=\columnwidth]{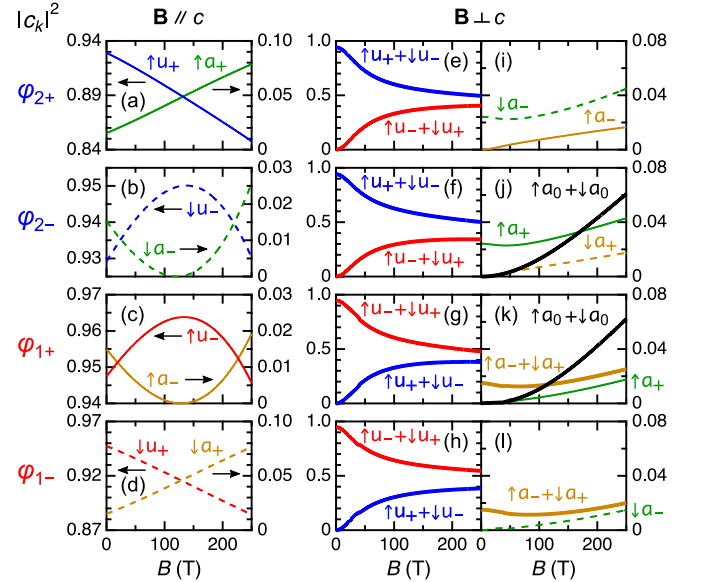}
\caption{Square of the coefficient $|c_{k}|^{2}$ for each of the $\Ket{({t}_{2}^{3}~^{2}E){\pm\frac{1}{2}u_{\pm}}}$ and $\Ket{({t}_{2}^{3}~^{2}T_{1}){\pm\frac{1}{2}a_{\pm, 0}}}$ components in the wave function of four Zeeman levels related to the R lines with the magnetic field up to 250~T. Data for ${\bf B} \parallel c$ and ${\bf B} \perp c$ are shown in (a)--(d) and (e)--(l), respectively. The figures are arranged in the same raw for each of the four levels in order of $\ket{\varphi_{2+}}$, $\ket{\varphi_{2-}}$, $\ket{\varphi_{1+}}$, and $\ket{\varphi_{1-}}$ from the top. Arrows $\uparrow$ ($\downarrow$) represent the spin $M_{S}=+\frac{1}{2}$ ($-\frac{1}{2}$) of the bases. For clarity, some plots show the summation of two components that mainly contribute to the same Kramers doublet \cite{comment2}.}
\label{Wavefunction}
\end{figure}

\vspace{-0.3cm}
\section{\label{sec:level4}DISCUSSIONS}
\vspace{-0.2cm}

\subsection{Wave functions in magnetic fields}
\vspace{-0.2cm}

To clarify the origin of the AZ effect above 100~T, we discuss the wave functions of the four Zeeman levels splitting from the $^{2}E$ states.
The wave functions are expressed by a linear combination of the 120 bases in the $3d^{3}$ state, $\ket{\varphi_{r}}=\sum_{k}^{}{c_{k}\ket{(\alpha S\Gamma) M_{S}\gamma}}$, where $c_{k}$ is a complex number coefficient.
Our calculation reveals that $\ket{\varphi_{1\pm}}$ ($\ket{\varphi_{2\pm}}$) at zero field mainly consists of one base $\Ket{(t_{2}^{3}~^{2}E){\pm\frac{1}{2}u_{\mp}}}\mathopen{\left(\vphantom{\Ket{(t_{2}^{3}~^{2}E){\pm\frac{1}{2}u_{\pm}}}}\kern-\nulldelimiterspace\right.}\Ket{(t_{2}^{3}~^{2}E){\pm\frac{1}{2}u_{\pm}}}\mathclose{\left.\kern-\nulldelimiterspace\vphantom{\Ket{(t_{2}^{3}~^{2}E){\pm\frac{1}{2}u_{\pm}}}}\right)}$ with $|c_{k}|^{2}=0.947~(0.941)$ \cite{comment1}.
When a magnetic field is applied, the second excited states $^{2}T_{1}$ start to hybridize with the $^{2}E$ states in the first order via the orbital term of ${\mathcal H}_{\rm Z}$.
Note that the $^{2}T_{1}$ states are composed of three Kramers doublets, all of which are $\Delta=60\sim90$~meV away from the $^{2}E$ states \cite{comment2}.
The field dependences of $|c_{k}|^{2}$ for the $^{2}E$ and $^{2}T_{1}$ states up to 250~T are summarized in Fig. \ref{Wavefunction}, where only the crucial components are shown.
For ${\bf B} \parallel c$, as shown in Figs. \ref{Wavefunction}(a)--\ref{Wavefunction}(d), the main component of the $^{2}E$ states decreases in each of the four wave functions, associated with the increase in the contribution of the $^{2}T_{1}$ states with the same $M_{L}$ and $M_{S}$.
Meanwhile, their field-dependences in $\ket{\varphi_{2-}}$ and $\ket{\varphi_{1+}}$ are different from those in $\ket{\varphi_{2+}}$ and $\ket{\varphi_{1-}}$.
This is responsible for the difference in the field dependence of $g'$, exhibiting convex upward and downward behaviors, respectively [see Fig. \ref{Calculation}(c)].
In contrast, for ${\bf B} \perp c$, both the summation of the $\Ket{+\frac{1}{2}u_{+}}$ and $\Ket{-\frac{1}{2}u_{-}}$ components and that of the $\Ket{+\frac{1}{2}u_{-}}$ and $\Ket{-\frac{1}{2}u_{+}}$ components approach 0.5 toward high magnetic fields as shown in Figs. \ref{Wavefunction}(e)--\ref{Wavefunction}(h).
These features signal the tendency of the quantization of the spin along the field direction, i.e. the previously reported PB effect \cite{1963_Aoy}.
Besides, as shown in Figs. \ref{Wavefunction}(i)--\ref{Wavefunction}(l), we see that the $\Ket{\pm\frac{1}{2}a_{0}}$ components significantly increase only in $\ket{\varphi_{2-}}$ and $\ket{\varphi_{1+}}$.
They are related to the lowest energy level of the three Kramers doublets of the $^{2}T_{1}$ states.
Therefore, the hybridization of those components is responsible
 for the shift of $E_{2-}$ and $E_{1+}$ to lower energy than $E_{2+}$ and $E_{1-}$, respectively [see Fig. \ref{Calculation}(b)].

\vspace{-0.3cm}
\subsection{Crystal-field Paschen-Back effect}
\vspace{-0.2cm}

These detailed theoretical analyses enable us to get more insights about the Zeeman patterns in ultrahigh magnetic fields.
As previously reported, the Zeeman patterns of the R lines exhibit (almost) linear behaviors around 60~T.
This is because of the special energy relations among the initial level splittings and the Zeeman energy, $\delta<{\mathcal H}_{\rm Z}\ll\Delta$, where the hybridization between the first and second excited states $^{2}E$ and $^{2}T_{1}$ is negligible.
However, their hybridization is already important above 100~T, resulting in the nonlinear behaviors that are experimentally observable.
Indeed, our calculation suggests that $2 \sim 12$~\% of the $^{2}T_{1}$ components are contributing to the R lines at 250~T as seen form Figs. \ref{Wavefunction}(a)--\ref{Wavefunction}(d) and Figs. \ref{Wavefunction}(i)--\ref{Wavefunction}(l).
In the field region of $10^{3}$~T, where $\Delta < {\mathcal H}_{\rm Z} \ll 10Dq$ is achieved, the $^{2}E$ and $^{2}T_{1}$ states are completely mixed.
Importantly, both of them do not directly interact with the higher excited states $^{4}T_{2}$, $^{2}T_{2}$, and $^{4}T_{1}$ shown in Fig. \ref{Introduction}(a) via the orbital term of ${\mathcal H}_{\rm Z}$.
Thus, the Zeeman patterns of these $^{2}E$ and $^{2}T_{1}$ states would approach the linear behaviors $\mu_{\rm B}(kM_{L}+2M_{S})B$ with $k=0.72$, $M_{L}=2, 1, 0, -1, -2$, and $M_{S}=1/2, -1/2$ (For detailed energy diagrams, see Fig. \ref{High_field} in Appendix B).
It is noteworthy that on further increasing the magnetic field much higher than $10^{4}$~T, where ${\mathcal H}_{\rm Z} \gg 10Dq$ is achieved, the Zeeman patterns of all the 120 levels would be finally quantized with $M_{L}+2M_{S}=6, 5, 4, \cdots, -5, -6$ (Here, we assume $k=k'=1$ because the covalent bonds could be broken in such high magnetic fields).

The Zeeman patterns of ruby discussed above can be regarded as a kind of PB effect.
However, to the best of our knowledge, such PB effect has not be proposed for describing the crossover from the AZ effect to the NZ effect under the crystal-field splitting, possibly due to the lack of proper situation.
The energy scale of the crystal field as well as the complexity of the level splittings make it challenging to clearly observe the {\it crystal-field} PB effect.
Note that the magnetic field accessible by the current technology is at most $10^{2}\sim10^{3}$~T \cite{2003_Miu, 2018_Nak}.
Therefore, our experimental observation of the onset of the {\it crystal-field} PB effect above 100 T owes to the accidental crystal-field splitting manner in ruby, i.e. the rather small energy gap $\Delta$ between the $^{2}E$ and $^{2}T_{1}$ states of the 3$d^{3}$ multiplets.

\vspace{-0.3cm}
\section{\label{sec:level5}CONCLUSION}
\vspace{-0.2cm}

In conclusion, the AZ effect was observed for the R lines of ruby above 100~T.
The crystal-field multiplet theory with several empirical parameters successfully reproduces the experimental Zeeman patterns up to 230~T, proving that the hybridization with the second excited states are responsible for their nonlinear behaviors.
Notably, the observed Zeeman patterns for ${\bf B} \perp c$ signal the crossover from one to another PB effect, characterized by the renormalization of the good quantum number from only $M_{S}$ to both $M_{L}$ and $M_{S}$.
The present work offers a new kind of PB effect, which is distinct from the conventional PB effects describing atomic energy levels, and it could be universally encountered in crystal if the conditions are met.

\vspace{-0.3cm}
\section*{ACKNOWLEDGMENTS}
\vspace{-0.2cm}

This work was partly supported by the JSPS KAKENHI Grants-In-Aid for Scientific Research (No. 18H01163, No. 19K23421, and No. 20J10988) and carried out in the science camp course for undergraduate students at ISSP.
M.G. was supported by the JSPS through a Grant-in-Aid for JSPS Fellows.
The authors thank T. Anan, Y. Ebihara, K. Kawauchi, and R. Nagashima for joining the science camp course.
M.G. and T.N. also thank M. Hagiwara and S. Takeyama for fruitful discussions.

\vspace{-0.3cm}
\section*{APPENDIX A: RELATIVE PEAK INTENSITIES}
\vspace{-0.2cm}

\begin{figure}[t]
\centering
\includegraphics[width=\linewidth]{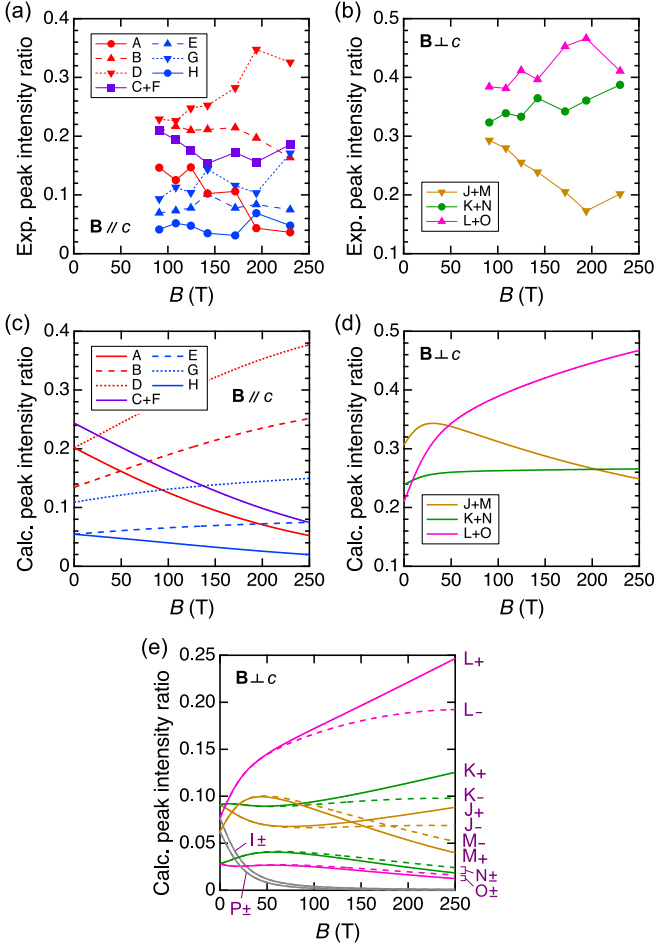}
\caption{(a)--(d) Magnetic-field dependences of the relative peak intensities obtained from the multi-Lorentzian fits for the experimental results [(a) and (b)] and the calculation based on Eq. \ref{eq:Intensity} [(c) and (d)]. In the calculation, the temperature is fixed to $T=200$~K to derive the Boltzmann distribution within the $^{2}E$ state. Plots of the calculated intensity ratios for all of the individual peaks for ${\bf B} \perp c$ geometry, which are decomposed from the plots in (d).} 
\label{Intensity}
\end{figure}

The peak assignments in Fig. \ref{Experiment} are confirmed by analyzing the peak intensities of the PL spectra.
In Figs. \ref{Intensity}(a) and \ref{Intensity}(b), we plot the field dependence of the relative peak intensities for ${\bf B} \parallel c$ and ${\bf B} \perp c$, respectively, which are obtained from the multi-Lorentzian fits of the experimental PL spectra.
Below, the relative peak intensities are calculated based on the crystal-field theory and compared with the experimental results.

For simplicity, we assume that the Zeeman levels of the excited state $^{2}E$ are purely composed of the orbital functions $u_{\pm}$ in the framework of the effective Hamiltonian \cite{1958_Sug_1}.
Thus, these wave functions can be expressed as $\ket{\varphi_{1\pm}}=\Ket{\pm\frac{1}{2}u_{\mp}}$ and $\ket{\varphi_{2\pm}}=\Ket{\pm\frac{1}{2}u_{\pm}}$ for ${\bf B} \parallel c$, and $\ket{\varphi_{1\pm}}=\cos\theta\Ket{\pm\frac{1}{2}u_{\mp}}-\sin\theta\Ket{\pm\frac{1}{2}u_{\pm}}$ and $\ket{\varphi_{2\pm}}=\sin\theta\Ket{\pm\frac{1}{2}u_{\mp}}+\cos\theta\Ket{\pm\frac{1}{2}u_{\pm}}$ for ${\bf B} \perp c$, where $\theta=\frac{1}{2}\arctan(g_{0}\mu_{\rm B}B/\delta)$.
The wave functions of the Zeeman levels of the ground state $^{4}A_{2}$ can also be expressed as $\ket{\phi_{\pm3/2}}=\Ket{\pm\frac{3}{2}e_{2}}$ and $\ket{\phi_{\pm1/2}}=\Ket{\pm\frac{1}{2}e_{2}}$, where the spin $M_{S}$ is defined so as to be quantized along the magnetic-field direction.
Then, the relative peak intensity corresponding to the transition from the Zeeman level with the $\ket{\varphi_{r}}$ component in the initial $^{2}E$ state to that with the $\ket{\phi_{s}}$ component in the final $^{4}A_{2}$ state can be derived from the following formula:
\begin{equation}
\label{eq:Intensity}
I[{^{2}E}\ket{\varphi_{r}} \rightarrow {^{4}A}_{2}\ket{\phi_{s}}]=P[{^{2}E}\ket{\varphi_{r}}] \times W[{^{2}E}\ket{\varphi_{r}} \rightarrow {^{4}A_{2}}\ket{\phi_{s}}],
\vspace{+0.2cm}
\end{equation}
where $P[{^{2}E}\ket{\varphi_{r}}]$ is the population ratio of electrons in the Zeeman level with the $\ket{\varphi_{r}}$ component and $W[{^{2}E}\ket{\varphi_{r}} \rightarrow {^{4}A_{2}}\ket{\phi_{s}}]$ is the transition probability.

The population ratio $P$ is estimated by assuming the Boltzmann distribution of electrons among the Zeeman levels of the $^{2}E$ state.
Note that the measurement temperature in this work could be high enough to achieve the fast spin relaxation, yielding the nearly thermal equilibrium condition during the pulse.
Accordingly, $P$ is expressed as
\begin{equation}
\label{eq:Population}
P[{^{2}E}\ket{\varphi_{r}}]=\frac{e^{-\beta E_{r}}}{\displaystyle \sum_{r'=1\pm,2\pm}^{} e^{-\beta E_{r'}}},
\end{equation}
where $E_{r}$ is the energy of the Zeeman level with the $\ket{\varphi_{r}}$ component and $\beta=1/k_{\rm B}T$, where $k_{\rm B}$ is the Boltzmann's constant.
Here, we adopt the value of $E_{r}$ obtained from the crystal-field multiplet theory as shown in Figs. \ref{Calculation}(a) and \ref{Calculation}(b) (not from the effective Hamiltonian) with the fixed temperature $T=200$~K.

As for the transition probability $W$, the transition from the $^{2}E$ to the $^{4}A_{2}$ state is spin-prohibited.
Thus, we have to consider the second-order perturbation via the intermediate $^{4}T_{2}$ state through the spin-orbit interaction \cite{1958_Sug_1}. 
Accordingly, the dipole strength for the transition from the $\ket{\varphi_{r}}$ component in the $^{2}E$ state to the $\ket{\phi_{s}}$ component in the $^{4}A_{2}$ state is given by the following formula:
\begin{equation}
\begin{split}
\label{eq:Transition}
&W[{^{2}E}\ket{\varphi_{r}} \rightarrow {^{4}A_{2}}\ket{\phi_{s}}] \propto \left| \bra{(^{4}A_{2}) \phi_{s}}{\overline {\mathbf P}}\ket{(^{2}E) \varphi_{r}} \right|^{2}\\
&=\left| \sum_{m}^{} \frac{\bra{(^{4}A_{2}) \phi_{s}}{\overline {\mathbf P}}\ket{(^{4}T_{2}) \psi_{m}}\bra{(^{4}T_{2}) \psi_{m}}{{\mathcal H}_{\rm SO}}\ket{(^{2}E) \varphi_{r}}}{E(^{2}E)-E(^{4}T_{2})} \right|^{2},
\end{split}
\end{equation}
where $\ket{\psi_{m}}$ is the component of the intermediate $^{4}T_{2}$ state and $E(^{2}E)-E(^{4}T_{2})$ is the energy gap between the $^{2}E$ and $^{4}T_{2}$ states.
If we neglect the effect of the phonon assisted transitions, the values of $\left| \bra{(^{4}A_{2}) \phi_{s}}{\overline {\mathbf P}}\ket{(^{2}E) \varphi_{r}} \right|^{2}$ can be obtained as Tables~\ref{tab:W_1} and \ref{tab:W_2} for ${\bf B} \parallel c$ and ${\bf B} \perp c$, respectively \cite{1958_Sug_1, 1963_Aoy}.
Here, $\sigma$ and $\pi$ are defined as follows (spin terms are omitted):

\vspace{-0.5cm}
\begin{equation}
\label{eq:sigma}
\sigma=\frac{4}{3}\zeta'^{2}\left| \frac{\bra{(^{4}A_{2})e_{2}}{\overline {\mathbf P}}\ket{(^{4}T_{2})x_{\pm}}}{E(^{2}E)-E(^{4}T_{2})} \right|^{2},
\end{equation}
\vspace{-0.3cm}
\begin{equation}
\label{eq:pi}
\pi=\frac{4}{3}\zeta'^{2}\left| \frac{\bra{(^{4}A_{2})e_{2}}{\overline {\mathbf P}}\ket{(^{4}T_{2})x_{0}}}{E(^{2}E)-E(^{4}T_{2})} \right|^{2}.
\end{equation}

In our experimental setup, only the $\sigma$ component (${\bf E} \perp c$) should be detected in the PL spectra for ${\bf B} \parallel c$ geometry, whereas both the $\sigma$ and $\pi$ components are expected to coexist for ${\bf B} \perp c$ geometry.
Such an effect might be reflected on the difference in the intensity distribution of the R lines at zero field between the ${\bf B} \parallel c$ and ${\bf B} \perp c$ data; the relative intensity of the R$_{1}$ line is a bit weaker in the ${\bf B} \perp c$ data [see Fig. \ref{Experiment}(a)].
Note that the values of $\left| \bra{(^{4}A_{2}) \phi_{s}}{\overline {\mathbf P}}\ket{(^{2}E) \varphi_{r}} \right|^{2}$ for the R$_{1}$ and R$_{2}$ lines at zero field are respectively $(5/3)\sigma$ and $\sigma$ for ${\bf E} \perp c$, whereas $(1/3)\pi$ and $\pi$ for ${\bf E} \parallel c$.
By considering the Boltzmann distribution within the $^{2}E$ states as well as the above information, we estimate the experimental ratio of the values of $\sigma$ and $\pi$ as $\sigma \sim 7\pi$. 

From these preparations, the magnetic-field dependence of the peak intensity ratios for ${\bf B} \parallel c$ and ${\bf B} \perp c$ are calculated as Figs. \ref{Intensity}(c) and \ref{Intensity}(d), respectively, where several peak intensities are combined in order to compare with the experimental plots shown in Figs. \ref{Intensity}(a) and \ref{Intensity}(b).
Here, the summation of all the peak intensities are normalized to 1 at each magnetic field.

For ${\bf B} \parallel c$, the changes in the relative peak intensities up to 250~T mainly originate from the increased population of the lowest Zeeman level of the $^{2}E$ state because the transition probabilities $W$ remain constant in this field range.
Therefore, the relative peak intensities corresponding to the transitions from the Zeeman level with the $\ket{\varphi_{1-}}$ or $\ket{\varphi_{2-}}$ component, i.e. the peaks B, D, E, and G, are expected to get stronger as the magnetic field increases, which is roughly consistent with the experimental results as shown in Fig. \ref{Intensity}(a).

For ${\bf B} \perp c$, in contrast, the field dependences of the relative peak intensities are more complicated because the hybridizations of two components in the Zeeman levels of the $^{2}E$ state bring about the changes in the transition probabilities $W$ as shown in Table \ref{tab:W_2}.
Since it is challenging to make the perfect peak assignments on the experimental PL spectra, we plot the relative intensities for three groups of the peaks (J+M, K+N, and L+O) in Fig. \ref{Intensity}(b) to avoid the ambiguous plots.
Indeed, our crystal-field multiplet theory predicts that each of these three groups above 100~T are composed of four peaks.
However, it is noteworthy that two peaks in the L+O group become strong toward high magnetic fields, as is clearly seen in Fig. \ref{Experiment}(a).
This tendency can be understood from the present calculation of the peak intensities.
Figure \ref{Intensity}(e) shows the calculated intensities for all of the individual peaks with the magnetic field ${\bf B} \perp c$, indicating that the peaks L$\pm$ get much stronger than the peaks O$\pm$.
Hence, we conclude that the optical transitions ${^{2}E} \ket{\varphi_{1\pm}} \rightarrow {^{4}A_{2}} \ket{\phi_{-3/2}}$ are responsible for two intense peaks observed in the high energy side ($\sim 20$~meV) at 230~T.
In addition, the broad PL spectra observed in the low energy side ($\sim -30$~meV) at high magnetic fields could be attributed to the almost equal intensities among the four peaks J$\pm$ and M$\pm$ as shown in Fig. \ref{Intensity}(e).

\onecolumngrid

\begin{table}[h]
\renewcommand{\arraystretch}{1.2}
\caption{The values of $\left| \bra{(^{4}A_{2}) \phi_{s}}{\overline {\mathbf P}}\ket{(^{2}E) \varphi_{r}} \right|^{2}$ for ${\bf B} \parallel c$ geometry \cite{1958_Sug_1}.}
\begin{tabular}{c|cccc|cccc} \hline\hline
 ~~Polarization~~ & \multicolumn{4}{c|}{${\bf E} \perp c$} & \multicolumn{4}{c}{${\bf E} \parallel c$} \\ \hline
 $^{4}A_{2}$ \verb|\| $^{2}E$ & ~~~$\varphi_{1+}$~~~ & ~~~$\varphi_{1-}$~~~ & ~~~$\varphi_{2+}$~~~ & ~~~$\varphi_{2-}$~~~ & ~~~$\varphi_{1+}$~~~ & ~~~$\varphi_{1-}$~~~ & ~~~$\varphi_{2+}$~~~ & ~~~$\varphi_{2-}$~~~ \\ \hline
$\phi_{+3/2}$ & $\sigma/2$ & & & & & & $\pi/2$ & \\
$\phi_{+1/2}$ & $\sigma/3$ & & $\sigma/3$ & $\sigma/6$ & & $\pi/6$ & & \\
$\phi_{-1/2}$ & & $\sigma/3$ & $\sigma/6$ & $\sigma/3$ & $\pi/6$ & & & \\
$\phi_{-3/2}$ & & $\sigma/2$ & & & & & & $\pi/2$ \\ \hline\hline
\end{tabular}
\label{tab:W_1}
\vspace{+0.2cm}
\renewcommand{\arraystretch}{1.2}
\caption{The values of $\left| \bra{(^{4}A_{2}) \phi_{s}}{\overline {\mathbf P}}\ket{(^{2}E) \varphi_{r}} \right|^{2}$ for ${\bf B} \perp c$ geometry \cite{1963_Aoy}.}
\begin{tabular}{c|cc|cc} \hline\hline
 ~~Polarization~~ & \multicolumn{2}{c|}{${\bf E} \perp c$} & \multicolumn{2}{c}{${\bf E} \parallel c$} \\ \hline
 $^{4}A_{2}$ \verb|\| $^{2}E$ & ~~~$\varphi_{1+}+\varphi_{1-}$~~~ & ~~~$\varphi_{2+}+\varphi_{2-}$~~~ & ~~~$\varphi_{1+}+\varphi_{1-}$~~~ & ~~~$\varphi_{2+}+\varphi_{2-}$~~~ \\ \hline
$\phi_{+3/2}$ & ~$3(\cos\theta-\sin\theta)^{2}\sigma$~ & ~$3(\cos\theta+\sin\theta)^{2}\sigma$~ & ~~$(\cos\theta-\sin\theta)^{2}\pi$~~ & ~~$(\cos\theta+\sin\theta)^{2}\pi$~~ \\
 \raisebox{0.5em}{$\phi_{+1/2}$} & \bettershortstack[c]{0.4}{1.2}{~$[2(\cos\theta+\sin\theta)^{2}$ \\ $+(3\cos\theta-\sin\theta)^{2}]\sigma/3$~} & \bettershortstack[c]{0.4}{1.2}{~$[2(\cos\theta-\sin\theta)^{2}$ \\ $+(\cos\theta+3\sin\theta)^{2}]\sigma/3$~} & \raisebox{0.5em}{~~$(\cos\theta-3\sin\theta)^{2}\pi/3$~~} & \raisebox{0.5em}{~~$(3\cos\theta+\sin\theta)^{2}\pi/3$~~} \\
 \raisebox{0.5em}{$\phi_{-1/2}$} & \bettershortstack[c]{0.4}{1.2}{~$[2(\cos\theta-\sin\theta)^{2}$ \\ $+(3\cos\theta+\sin\theta)^{2}]\sigma/3$~} & \bettershortstack[c]{0.4}{1.2}{~$[2(\cos\theta+\sin\theta)^{2}$ \\ $+(\cos\theta-3\sin\theta)^{2}]\sigma/3$~} & \raisebox{0.5em}{~~$(\cos\theta+3\sin\theta)^{2}\pi/3$~~} & \raisebox{0.5em}{~~$(3\cos\theta-\sin\theta)^{2}\pi/3$~~} \\
$\phi_{-3/2}$ & ~$3(\cos\theta+\sin\theta)^{2}\sigma$~ & ~$3(\cos\theta-\sin\theta)^{2}\sigma$~ & ~~$(\cos\theta+\sin\theta)^{2}\pi$~~ & ~~$(\cos\theta-\sin\theta)^{2}\pi$~~ \\ \hline\hline
\end{tabular}
\vspace{+0.7cm}
\label{tab:W_2}
\end{table}

\twocolumngrid

\vspace{-0.3cm}
\section*{APPENDIX B: THEORETICAL ENERGY DIAGRAMS TOWARD THE HIGH-FIELD LIMIT}
\vspace{-0.2cm}

\begin{figure*}[t]
\centering
\includegraphics[width=\linewidth]{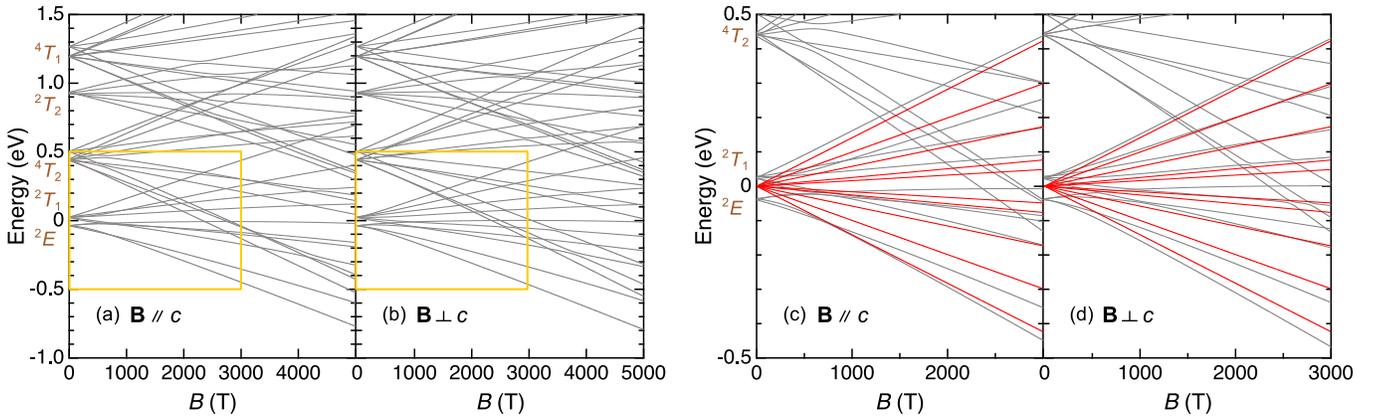}
\caption{(a)(b) Calculated energy diagrams with the magnetic field up to 5000~T for (a) ${\bf B} \parallel c$ and (b) ${\bf B} \perp c$ geometries. (c)(d) Enlarged view of the Zeeman patterns focusing on the $^{2}E$ and $^{2}T_{1}$ states. They correspond to the region surrounded by yellow squares in (a) and (b). Expected asymptotic lines, $E=\mu_{\rm B}(kM_{L}+2M_{S})B$, are displayed in red.}
\label{High_field}
\end{figure*}

The calculated energy diagrams extended up to 5000~T are shown in Figs. \ref{High_field}(a) and \ref{High_field}(b) for ${\bf B} \parallel c$ and ${\bf B} \perp c$ geometries, respectively.
The parameters are given in the main text.
Here, the energy is offset so that the mean energy of 10 levels originating from the first and second excited states $^{2}E$ and $^{2}T_{1}$ at zero field becomes 0.

As pointed out in the main text, the hybridization between the $^{2}E$ and $^{2}T_{1}$ states occurs above 100~T, yielding the AZ effect in the field region around $100 \sim 1000$~T.
On the other hand, once the Zeeman energy exceeds the initial splitting between the $^{2}E$ and $^{2}T_{1}$ states, their Zeeman patterns would asymptotically approach the linear behaviors, i.e. the NZ effect.
In Figs. \ref{High_field}(c) and \ref{High_field}(d), we show the enlarged views of their Zeeman patterns in together with the linear lines, $E=\mu_{\rm B}(kM_{L}+2M_{S})B$, where $kM_{L}+2M_{S}=\pm2.44, \pm1.72, \pm1, \pm0.44, \pm0.28$.
For both geometries (especially, for ${\bf B} \perp c$), the calculated Zeeman patterns are found to approach the expected linear lines above 1000~T, confirming the reconstruction of the good quantum number of the orbital angular momentum as $M_{L}=2, 1, 0, -1, -2$.
Note that some discrepancies come from the higher-order interactions with the upper excited states, resulting in the $g$-shifts for the individual lines or the global energy shift to the lower energy side.


\begin{thebibliography}{99}
\bibitem{comment1} These values are different from those of the intercepts shown in Figs. \ref{Wavefunction}(a)--\ref{Wavefunction}(h). This is  because $|c_{k}|^{2}$ of each component changes discontinuously due to the lifting of degeneracy when a magnetic field is applied.
\bibitem{comment2} The energy gap between the first and second excited states, $^{2}E$ and $^{2}T_{1}$, is attributed to the single-ion Coulomb interactions in the framework of the crystal-field theory. Three Kramers doublets originating from the $^{2}T_{1}$ states are represented as 2$\overline A$, $\overline E_{a}$, and $\overline E_{b}$. In ruby, the main components contributing to each of the doublet are as follows: $\Ket{+\frac{1}{2}a_{+}}$ and $\Ket{-\frac{1}{2}a_{-}}$ for 2$\overline A$, $\Ket{+\frac{1}{2}a_{-}}$ and $\Ket{-\frac{1}{2}a_{+}}$ for $\overline E_{a}$, and $\Ket{+\frac{1}{2}a_{0}}$ and $\Ket{-\frac{1}{2}a_{0}}$ for $\overline E_{b}$.
\end{thebibliography}
\end{document}